\documentclass[10pt,a4paper,superscriptaddress,aps,prd,nofootinbib,notitlepage]{revtex4}
\pdfoutput=1
\usepackage{lmodern}
\usepackage{verbatim}

\usepackage[T1]{fontenc}
\usepackage[utf8]{inputenc}
\usepackage{color}
\usepackage{amsmath}
\usepackage{amssymb}
\usepackage{graphicx}
\usepackage{esint}
\usepackage[unicode=true,bookmarks=true,bookmarksnumbered=false,bookmarksopen=false, breaklinks=false,pdfborder={0 0 1},backref=false,colorlinks=true] {hyperref}

\begin{document}

\title{Fisher matrix for multiple tracers: model independent constraints on the redshift distortion parameter}
\author{L. Raul Abramo}
\affiliation{Departamento de F\'{\i}sica Matem\'atica, Instituto de F\'{\i}sica, Universidade de S\~ao Paulo, Rua do Mat\~ao 
1371, CEP 05508-090, S\~ao Paulo, Brazil}
\author{Luca Amendola}
\affiliation{ITP, Ruprecht-Karls-Universität Heidelberg, Philosophenweg 16, 69120
Heidelberg, Germany}
\date{\today}

\begin{abstract}
We show how to obtain  constraints on $\beta=f/b$, the ratio of the matter growth rate and the bias that quantifies the linear redshift-space distortions, that are independent of the cosmological model, using multiple tracers of large-scale structure. 
For a single tracer the uncertainties on $\beta$ are constrained by the uncertainties in the amplitude and shape of the power spectrum, which is limited by  cosmic variance. However, for two or more tracers this limit does not apply, since taking the ratio of power spectra cosmic variance cancels out, and in the linear (Kaiser) approximation one measures directly the quantity $(1+ \beta_1 \mu^2)^2/(1+ \beta_2 \mu^2)^2$, where $\mu$ is the angle of a given mode with the line of sight. We provide analytic formulae for the Fisher matrix for one and two tracers, 
and quantify the signal-to-noise ratio needed to make effective use of the multiple-tracer technique. We also forecast the errors on $\beta$ for a survey like Euclid.
\end{abstract}

\maketitle

\section{Introduction}

At the scales at which the cosmological fluctuations are well within the linear regime, and at small redshifts, all the information on the cosmological model is contained in very few quantities, such as the Hubble function, the power spectrum of the galaxy density field and the power spectrum of the weak-lensing shear. A desirable goal of large scale observations is to derive from this information a measurement of quantities of cosmological interest without first assuming a particular model, e.g. $\Lambda$CDM. One such quantity is the redshift distortion parameter:
\begin{equation}
\label{Eq:beta}
\beta=\frac{f}{b} \; ,
\end{equation}
where $f=d\log\delta_{m}/{d\log a}$ is the matter growth rate [$a(t)$ is the cosmological scale factor] and $b=\delta_{g}/\delta_{m}$ is the linear galaxy bias. This quantity enters the linear galaxy power spectrum in the so-called Kaiser term \cite{kaiser_clustering_1987}, which converts the real-space spectrum $P^{(x)}(k) = b^2 P_m (k)$ ($P_m$ is the matter power spectrum) into its observed, redshift-space version, $P^{(z)}(k,\mu) = (1+ \beta \mu^2)^2 P^{(x)} (k)$, where the angle of the Fourier mode $\vec{k}$ with the line of sight $\hat{r}$ is encoded in $\mu = \hat{r} \cdot \hat{k}$ --- \cite{1977ApJ...212L...3S}, see \cite{Hamilton:1998} for a review. The quantity $\beta$  is also needed to define the statistics $E_{g}$ that helps to discriminate between different cosmological models -- see, e.g., \cite{Zhang2007,Leonard2015}.

If the underlying density field is sampled only by one kind of tracer (galaxies, QSOs, Lyman-$\alpha$ systems, etc.), the redshift-space distortion factor is degenerate with the amplitude of the power spectrum, which in turn is limited by cosmic variance.
Hence, for a single tracer the uncertainty on $\beta$, as well as many other interesting effects \cite{2015ApJ...814..145A}, will also be limited by cosmic variance, even in the limit of infinite signal-to-noise.

A possible way to overcome this limitation has been proposed in \cite{Seljak:2009,McDonald:2009}: the idea consists in identifying two (or more) tracers that sample the density field with number densities $n_{1,2}$ and with biases $b_{1,2}$, and measuring the ratios of the spectra of the two tracers. In this case it has been shown that ratios of spectra can be measured to arbitrarily high precision if the signal-to-noise ratio (SNR) is arbitrarily high -- see also, e.g., \cite{2009MNRAS.397.1348W,Hamaus:2010im,GilMarin:2010vf,Bernstein:2011ju,Abramo:2013}. 
This cancellation of cosmic variance can be used to improve measurements of bias and growth rate \cite{2009MNRAS.397.1348W,GilMarin:2010vf,Hamaus:2012ap,Koda:2013eya,Abramo:2013,Alarcon:2016bkr}, as well as primordial non-Gaussianities and horizon-size effects \cite{2011PhRvD..84h3509H,Ferramacho:2014pua,Ferraro:2014jba,Fonseca:2015laa,2015PhRvD..92f3525A,2016JCAP...08..021B,2017PhRvD..96l3535A}, and has been widely studied both in the context of galaxy surveys (e.g., \cite{2013MNRAS.436.3089B,2014arXiv1403.5237B}) as well as radio/21cm surveys \cite{2014PhRvD..90h3520Y,Ferramacho:2014pua,Fonseca:2015laa,Witzemann:2018cdx}.
Since the signal is the power spectrum $P_i$, and in the usual case of Poisson statistics for the counts of the tracers shot noise is given by $1/\bar{n}_i$ (where $\bar{n}_i$ is the mean number count of the tracer $i$), increasing the SNR=$\bar{n}_i P_i$ means observing larger numbers of the tracers. Of course, if $b_{1}=b_{2}$ the ratio of the two spectra is unity, and there is no gain with respect to a single-tracer survey.

The goal of this paper is to derive general estimates of the uncertainties on $\beta_{1,2}$ without assuming any particular cosmological model. From $\beta_{1,2}$ one can obviously obtain the bias ratio $b_1/b_2$. Notice, however, that without further independent information on $b_{1,2}$ nothing can be said about the growth rate $f$ itself.

We use the Fisher matrix formalism to obtain the errors for the power spectra and the parameters $\beta_{1,2}$ for two tracers by integrating out, in an essentially analytical way, the anisotropic clustering due to redshift-space distortions. This extends and clarifies several previous results: e.g., Ref. \cite{McDonald:2009} only considers modes along and across the line-of-sight; 
Ref. \cite{Hamaus:2012ap} computed analytically a Fisher matrix and employed N-body simulations to demonstrate the gain from splitting a survey into different biased tracers of large-scale structure;
Ref. \cite{Koda:2013eya} studies how the combination of a galaxy survey and a peculiar velocity 
survey can constrain $\beta$ under some assumptions which we can relax; 
and Ref. \cite{Alarcon:2016bkr} obtain independent constraints on $\beta$, 
but only after constraining the power spectra.
Ref. \cite{2013MNRAS.436.3089B} also performed a Fisher forecast on $f(z)$ and $b_{1,2}$ for two tracers, assuming standard power spectra and $k$-independent biases and growth rate.
Finally, we point out that the GAMA survey derived improved constraints on 
the matter growth rate by splitting their galaxies into luminosity and color classes 
\cite{2013MNRAS.436.3089B}, obtaining gains of $\sim$ 10 -- 20 \% compared with the 
single-tracer analysis. We estimated the gain in the model-independent determination of $\beta$ and the power spectrum using same number densities and biases of the tracers in the GAMA survey, and we found an improvement of $\sim$ 15 \% in the two-tracer case as compared with the single tracer -- although it should be stressed that the GAMA constraints were obtained under the standard assumptions about the cosmological model, whereas our constraints are independent of the model.

Our results will be expressed in terms of the Fisher matrix per unit volume of phase-space (let us call this the \emph{ Fisher matrix density}), i.e. for a given bin in $k$ and $z$. In this way, our results are not only model independent, but also survey independent. For any particular survey, one has to multiply the matrix by the phase-space volume element to obtain the final errors. We will give an example based on the Euclid survey \cite{Laureijs:2011gra,Euclidmission}, but our results are relevant for any deep survey with large number densities of tracers, and spanning large cosmological volumes, such as, e.g., J-PAS \cite{2014arXiv1403.5237B} and SPHEREx \cite{2016arXiv160607039D}.

It is worth stressing the importance of deriving results that are independent of a specific model. Whenever the estimation of a quantity depends on assuming a model or a class of model defined by some set of parameters, e.g. $\Lambda$CDM, then that estimation  cannot be employed in any other context except for that model. For instance, a measurement of $\beta$ that assumes $\Lambda$CDM cannot be employed in the statistics $E_g$ to test any other model, and therefore preempts its validity as a test of gravity. The statistics can still be used as a null test, i.e. to  reject models within the chosen class, but not to gauge their merits with respect to models outside the class. In statistical language, a quantity that is measured assuming a model can be used to produce frequentist $p$-value hypothesis tests, or nested Bayesian model selection within the parameters of the same class of models, but not general Bayesian model selection.

Being model-independent, our results are conservative: our only assumption is that we are on linear scales, where the Kaiser approximation is valid -- however, the same conclusions should apply if we abandon the flat-sky (distant observer) limit \cite{2011PhRvD..84d3516C,Yoo:2012se,Bonvin:2011,Bertacca:2012,2018MNRAS.481.1251C}. Given a specific model with a small number of parameters, one can sum the Fisher matrix over all $k-$ and $z-$bins and project the result onto the chosen parameters, obtaining stronger, but model-dependent, constraints. 

Typical values for the bias range from $b \lesssim 1$ for low-mass systems observed through Ly-$\alpha$ absorption or through the neutral H 21cm line (see, e.g., \cite{2015ApJ...803...21B}), all the way to $\sim 5-10$ for very massive halos at high redshifts \cite{2009ApJ...697.1634R}. Since the matter growth rate $f\sim 0.5 - 1$ in the interesting redshift range, $\beta$ is expected to assume values in the range 0.1 -- 2.0. It is interesting to note that  one could include, in addition to halos or galaxies, also voids, which have negative bias \cite{2018arXiv181204024C}.

\section{Fisher matrix for multiple tracers}

We can derive the Fisher matrix for two or more tracers (e.g., different kinds of galaxies, quasars, halos of different masses, etc.) starting directly from the density contrast in Fourier space, rather than from the power spectrum as sometimes done in the literature (e.g. \cite{1997PhRvL..79.3806T}). Although the result is the same, in standard cosmology it's the density contrast that is a Gaussian variable, not the power spectrum.

Let us consider two distinct Gaussian fields at some redshift $z$, with zero mean, sampled by distinct ``particles'' (e.g., two different galaxy populations). Let $\delta_{1,2}$ be their $k$-th Fourier coefficient and $\bar n_{1,2}$ their number density. For now, we do not have to assume any redshift distortion. We assume that the only non-zero correlations in a survey of volume $V$ are ($i,j=1,2$):
\begin{equation}
\label{Eq:Vdelta}
V\langle\delta_{i}\delta_{j}^{*}\rangle =
b_i b_j P_m + \frac{\delta_{ij}}{\bar{n}_i} =
\left\{ 
\begin{array}{ccc}
P_{i}N_{i} & {\,} & (i=j) \\
\sqrt{P_i P_j} & {\,} & (i\neq j) 
\end{array}
\right. \; ,
\end{equation}
where $\langle \delta_i\delta_i\rangle=0$, $P_i$ are the power spectra as a function of $z$ and $\vec{k}$, and $N_{i}=1+ 1/(\bar{n}_{i}P_{i})$ is the shot-noise term -- from now on the biases $b_{i}$ are absorbed in the power spectra.  
Notice that since we assume that the two discrete realizations of the field are formed by distinct particles, there
is no cross shot-noise. 
However, it should be stressed that halo exclusion effects and non-linear clustering introduce corrections to the usual Poissonian shot noise, which in reality may not be well represented by a diagonal matrix $\delta_{ij}/\bar{n}_i$ \cite{Hamaus:2010im,2013PhRvD..88h3507B}. 
These corrections have been used to improve forecasts \cite{Hamaus:2012ap}, and could also be applied to the model-independent approach that we present here.

From now on we assume that the spectra are expressed
in units of shot noise, so all the factors of 
$\bar{n}_{i} \to 1$, hence $\bar{n}_i P_i \to P_i$, which are henceforth adimensional. 
For any given $\vec k$, the random variables $x_a=\sqrt{V\bar n_i}\{\delta_{1},\delta_{1}^{*},\delta_{2},\delta_{2}^{*}\}$ are distributed as (sum over repeated indexes)
\begin{equation}
\label{Eq:likelihood}
L=\frac{1}{(2\pi)^{2}|C|^{1/2}}\exp\left(-\frac{1}{2}x_{a}C_{ab}^{-1}x_{b}\right) \; ,
\end{equation}
where the correlation matrix is:
\begin{equation}
\label{Eq:correlation}
C_{ab}=\langle x_{a}x_{b}\rangle=\left(\begin{array}{cccc}
0 & P_{1}N_{1} & 0 & \sqrt{P_{1}P_{2}}\\
P_{1}N_{1} & 0 & \sqrt{P_{1}P_{2}} & 0\\
0 & \sqrt{P_{1}P_{2}} & 0 & P_{2}N_{2}\\
\sqrt{P_{1}P_{2}} & 0 & P_{2}N_{2} & 0
\end{array}\right) \; .
\end{equation}
The Fisher matrix for a set of parameters $\theta_{\alpha}$, in a survey of volume $V$, is then
\begin{equation}
\label{Eq:Fishbar}
F_{\alpha\beta}=\frac12 \frac{4\pi k^{2}\Delta_{k}}{(2\pi)^{3}}V\bar{F}_{\alpha\beta}= V V_{k}\bar{F}_{\alpha\beta} \; ,
\end{equation}
where an extra factor of $1/2$ in the Fisher matrix accounts for the 
overcounting of the degrees of freedom in the likelihood since $\delta_{\vec k}^{*}=\delta_{-\vec k}$.
We include this factor in the effective volume in Fourier space, 
$V_{k}=\frac{4\pi k^{2}\Delta_{k}}{2(2\pi)^{3}}$, 
where $\Delta_k$ is the width of the bandpower (shell) $k$.
In the expression above $\bar{F}$ is the Fisher matrix per unit phase-space volume,
i.e., the Fisher matrix density\footnote{We point out that
the choice of dataset $x_a$ used here implies that the Fisher matrix density and phase volumes that are used in this paper differ by a factor of $1/2$ and $2$, respectively, from those used \cite{Abramo:2013,2016MNRAS.455.3871A}. The Fisher matrix, Eq. (\ref{Eq:Fishbar}), is of course identical.}:
\begin{equation}
\label{Eq:Fishbar2}
\bar{F}_{\alpha\beta}=\frac{1}{2}\frac{\partial C_{ab}}{\partial\theta_{\alpha}}C_{ad}^{-1}C_{bc}^{-1}\frac{\partial C_{cd}}{\partial\theta_{\beta}}
\end{equation}
If we take as parameters $\theta_{\alpha}=\{\log P_{1},\log P_{2}\}$, we obtain
\begin{equation}
\label{Eq:Fishbar3}
\bar{F}=\left(\begin{array}{cc}
\frac{\left(N_{1}-4\right)N_{2}+2N_{2}^{2}+1}{2\left(N_{1}N_{2}-1\right){}^{2}} & \frac{N_{1}\left(N_{2}-2\right)-2N_{2}+3}{2\left(N_{1}N_{2}-1\right){}^{2}}\\
\frac{N_{1}\left(N_{2}-2\right)-2N_{2}+3}{2\left(N_{1}N_{2}-1\right){}^{2}} & \frac{N_{1}\left(N_{2}-4\right)+2N_{1}^{2}+1}{2\left(N_{1}N_{2}-1\right){}^{2}}
\end{array}\right)
\end{equation}
which can be cast in a more elegant format:
\begin{equation}
\bar{F}=\frac{P_{1}P_{2}}{2\left(1+P\right){}^{2}}\left(\begin{array}{cc}
1+P+2R & 1-P\\
1-P & 1+P+\frac{2}{R}
\end{array}\right)\label{eq:trac1} \; ,
\end{equation}
where
\begin{align}
P & =P_{1}+P_{2}  \\
R & =\frac{P_{1}}{P_{2}}  \; .
\end{align}
The Fisher matrix above, in terms of the parameters $\theta=\{\log P_{1},\log P_{2}\}$, 
can be diagonalized by projecting onto the new variables $\Theta=\{\log P,\log R\}$. 
The Jacobian for this change of variables is
\begin{equation}
J=\frac{\partial \theta}{\partial \Theta} \; ,
\end{equation}
leading to:
\begin{equation}
\label{eq:f11}
F_{\Theta}=J^{T}FJ=\left(\begin{array}{cc}
\frac{P^{2}}{\left(P+1\right){}^{2}} & 0\\
0 & \frac{P_{1}P_{2}}{2\left(P+1\right)}
\end{array}\right)  \; .
\end{equation}
The marginalized relative error (from now on, all errors and variances are meant to be relative values) on $R$ is therefore 
\begin{equation}
\sigma_{R}^{2}=2 (VV_{k})^{-1}\frac{(1+P)}{P_{1}P_{2}}
\end{equation}
and decreases as $2/P_{2}\ll 1$ if $P_{1} \gg P_{2} \gg1$ (and vice-versa for $P_{2}$).

In the case of $N_t$ tracers, such that the degrees of freedom 
are $\theta=\{\log P_{1},\log P_{2}, \ldots, \log P_{N_t}\}$, the Fisher matrix density
can be written as \cite{Abramo:2013,2016MNRAS.455.3871A}:
\begin{equation}
\bar{F}_{ij} = \frac{1}{2} \frac{\delta_{ij} {P}_i {P} (1+{P}) + {P}_i {P}_j (1-{P})}{(1+{P})^2} \; ,
\label{Eq:MTFish}
\end{equation}
where $P = \sum_i^{N_t} P_i$. The diagonalized degrees of freedom in this case are
found by transforming from the ``Cartesian'' coordinates $\theta$ into hyper-spherical coordinates
$\Theta$ -- see \cite{Abramo:2013}. The errors on the new ``angle'' variables (which are simply ratios of spectra) 
scale in the same way as was found above for the ratio $R$ in the case of two tracers, i.e., 
their relative uncertainties can be arbitrarily small if the power spectra (in units of shot noise) have 
high enough amplitudes.

In order to obtain the actual errors for a given survey we must multiply the Fisher matrix density by 
the phase space volume factor.
For a typical scale of $\lambda=$100 Mpc$/h$ in a spherical (full sky) survey of radius $L$, 
one has, for a bandwidth $\Delta_{k}=2\pi(\Delta\lambda/\lambda^{2})\approx2\pi/L$:
\begin{equation}
\label{Eq:gamma}
\gamma^2=(VV_{k})^{-1}=\frac{3}{8\pi^{2}}\frac{\lambda^{3}}{L^{3}}\approx0.04\frac{\lambda^{3}}{L^{3}} \; .
\end{equation}
For $\lambda=$100 Mpc$/h$ and $L=1$ Gpc this factor amounts to $\approx 4 \cdot 10^{-5}$.
In the following we focus on the Fisher matrix density, so the errors
we derive should be multiplied by $\gamma$.
Assuming a survey like Euclid \citep{Laureijs:2011gra,Euclidmission} one can compute the $\gamma$ factors
for a given scale $k=2\pi/\lambda$  well within the linear regime, as reported in Table 
(\ref{tab:An-approximation-to}). Here,  $\Delta_k=2\pi/L$ in each bin has been estimated simply by using $L=V^{1/3}$, where $V$ is the comoving volume of the redshift bin. 

\begin{table}
\[
\begin{array}{cccccccc}
z_{1} & z_{2} & \bar{n}(\times10^{-3}) & \text{V}(\times10^{9}\text{Mpc}^{3}) & \bar{n}P(z,k_1)  & \gamma(k_1)  & \bar{n}P(z,k_2)  & \gamma(k_2)  \\
\hline

0.7 & 0.9 & 1.90 & 7.18 & 19.6 & 0.13 & 10.7 & 0.026 \\
0.9 & 1.1 & 1.71(1.46-0.86) & 9.02 & 14.8(12.6-7.45) & 0.12 & 8.07(6.89-4.07) & 0.024 \\
1.1 & 1.3 & 1.37(0.83-0.44) & 10.5 & 10.0(6.05-3.23) & 0.11 & 5.47(3.30-1.76) & 0.023 \\
1.3 & 1.5 & 0.99(0.43-0.23) & 11.6 & 6.19(2.71-1.42) & 0.11 & 3.38(1.48-0.77) & 0.022 \\
1.5 & 2.1 & 0.33 & 39.0 & 1.55 & 0.074 & 0.847 & 0.015 
\end{array}
\]
\caption{\label{tab:An-approximation-to}
Survey specifications for a   Euclid-like survey,
evaluated at scales $k_{1}=0.01$ $h$ Mpc$^{-1}$ and $k_{2}=0.05$
$h$ Mpc$^{-1}$. The densities $\bar n$ are taken from Ref. \cite{Euclidmission}, while the two values in parentheses are the more recent estimations of the galaxy density from \cite{Merson:2019vfr}, Table 2, WISP calibration and HiZELS calibration, respectively. The power spectrum is taken to be $\Lambda$CDM with Planck values \cite{PlanckCollaboration2015}, including non-linear correction.  }
\end{table}

\section{Model independence and statistical independence}

Before proceeding forward, we deem useful to spend a few words to clarify further the concept of model independence as employed in this paper. 

The fact that the Fisher matrix, Eq. (\ref{eq:f11}), is diagonal implies that the parameters that enter the total power spectrum are decoupled from those that enter  the ratio $R$. In this case, the estimation of $R$ is completely independent of any parametrization of $P$. This fact is at the core of the multiple-tracer method: although the error on $P$ reaches a limit for large SNR,  $R$ can be estimated to infinite precision for infinite SNR. However, as we will see later on, the general Fisher matrix for $\beta$ and $P$ when the Kaiser redshift distortion is included, is no longer diagonal, neither for the single- nor the multi-tracer case -- except in the infinite signal-to-noise ratio, as also shown in Ref. \cite{Hamaus:2012ap}. This means that the assumptions on $P$ (e.g., a $\Lambda$CDM spectrum) have an impact on the estimation and on the errors derived for $\beta$. Only by taking $P(k)$ itself as a parameter is model independence restored, since $P(k)$ is a directly observed quantity. 
One can make these considerations semantically more precise. Suppose we have only two unknown parameters to estimate, $A$ and $B$. If $A$ and/or $B$ can be directly estimated from observations,  we say they are model independent. In other words, a quantity is model independent only if it is a statistics. If, on the other hand, the estimation of $A$ (regardless of whether this is done directly from the data, or by first making assumptions such as, e.g., that $A$ is independent of $k$) is independent of the estimation of $B$, then we say that $A$ and $B$ are statistically independent. So two quantities are statistically independent if their Fisher matrix is diagonal. There is no {\it a priori} relation between the two concepts. Even if $A$ is statistically independent of $B$, one might still need to make assumptions in order to estimate it from the data, so it would not necessarily be a model-independent quantity. For instance $\beta$ is a model independent quantity, but $f$ is not, since it requires first the knowledge of $b$, or at least some parametrization.

As an obvious consequence of these definitions, the property of being model independent is not necessarily related to the multiple-tracer technique and does not require a diagonal Fisher matrix. As we show next, in fact, $\beta$ is a model-independent quantity also in the single-tracer case, where, just as for the multiple-tracer case with Kaiser correction, the Fisher matrix is in general not diagonal.

\section{Single-tracer case with Kaiser correction}

We now introduce the Kaiser correction $B=1+\beta\mu^{2}$.
Here, $\beta$ is supposed to be an unknown  function of $z$ and $k$, and whenever we refer to it we mean its value in a particular bin of $z,k$.
We begin with the example of a single tracer. 
We have the correlation matrix
\begin{equation}
C_{ab}=\left(\begin{array}{cc}
0 & B^{2}PN\\
B^{2}PN & 0
\end{array}\right) \; ,
\end{equation}
where now
\begin{equation}
N =1+\frac{1}{B ^{2}P } \; .
\end{equation}
Averaging the Fisher matrix density over the direction cosine $\mu$, we obtain
for the variables $\{\log P ,\log\beta\}$:
\begin{equation}
\bar{F}=\left(\begin{array}{cc}
\frac{1}{8}\left(\frac{2-2\left(\beta +1\right)P }{\left(P +1\right)\left(\left(\beta +1\right){}^{2}P +1\right)}+T_{1}+8\right) & \frac{1}{4}\left(\frac{2}{\left(\beta +1\right){}^{2}P +1}-T_{2}+8\right)\\
\frac{1}{4}\left(\frac{2}{\left(\beta +1\right){}^{2}P +1}-T_{2}+8\right) & \frac{\left(\beta +1\right)P +1}{\left(\beta +1\right){}^{2}P +1}+T_{3}+4
\end{array}\right) \; ,
\end{equation}
where $T_{i}=(z_{i}T)+(z_{i}T)^{*}$ and
\begin{equation}
T=\frac{\arctan\left(\frac{\sqrt{\beta P }}{\sqrt{i\sqrt{P }+P }}\right)}{\sqrt{\beta }\sqrt{i\sqrt{P }+P }} \; ,
\end{equation}
with
\begin{align}
z_{1} & =\frac{5-6i\sqrt{P }}{i+\sqrt{P }} \; , \\
z_{2} & =4\sqrt{P }+5i \; , \\
z_{3} & =-\frac{7}{2}\sqrt{P }+\frac{1}{2}i(-5+2P ) \; .
\end{align}
The marginalized \emph{relative} error on $\beta $ is
\begin{equation}
\sigma_{\beta }^{2}=\frac{\frac{2-2\left(\beta +1\right)P }{\left(P +1\right)\left(\left(\beta +1\right){}^{2}P +1\right)}+T_{1}+8}{\left(\frac{2-2\left(\beta +1\right)P }{\left(P +1\right)\left(\left(\beta +1\right){}^{2}P +1\right)}+T_{1}+8\right)\left(\frac{\left(\beta +1\right)P +1}{\left(\beta +1\right){}^{2}P +1}+T_{3}+4\right)-\frac{1}{2}\left(-\frac{2}{\left(\beta +1\right){}^{2}P +1}+T_{2}-8\right)^{2}}\label{eq:errbeta} \; .
\end{equation}
while for $P$ we have
\begin{equation}\label{eq:errp1}
  \sigma_{P }^{2}= \frac{\frac{(\beta +1) P+1}{(\beta +1)^2 P+1}+T_3+4}{\frac{1}{8} \left(\frac{2-2 (\beta +1) P}{(P+1) \left((\beta +1)^2
   P+1\right)}+T_1+8\right) \left(\frac{(\beta +1)
   P+1}{(\beta +1)^2 P+1}+T_3+4\right)-\frac{1}{16}
   \left(-\frac{2}{(\beta +1)^2 P+1}+T_2-8\right){}^2}
\end{equation}
   We find that $\sigma_{\beta }$ diverges for both large and small
$\beta $, and has a minimum at $\beta  \approx 12.31$ where $\sigma_{\beta } \approx 1.7$.
 For $P  \gg 1$ the result is independent of $P$:
\begin{equation}
\sigma_{\beta }^{2}=\frac{\beta \left(\beta +1\right)}{2\beta +2\left(\beta +1\right)(\sqrt{\beta }-2\tan^{-1}\sqrt{\beta})\tan^{-1}(\sqrt{\beta})} \; .
\label{eq:st-largep}
\end{equation}
We'll need also the relative error for $P$ in the same limit,
\begin{equation}
\sigma_{P }^{2}=\frac{\beta  \left(2 \beta +3\right)-3 \sqrt{\beta } \left(\beta +1\right) \tan
   ^{-1}\left(\sqrt{\beta }\right)}{\beta +\left(\beta +1\right) \tan
   ^{-1}\left(\sqrt{\beta }\right) \left(\sqrt{\beta }-2 \tan ^{-1}\left(\sqrt{\beta
   }\right)\right)} \; .
\label{eq:st-largepp}
\end{equation}
Notice that in order to derive $\mu$ and $k$ from observations based on redshifts and angles, one needs a background cosmological model, usually taken to be $\Lambda$CDM, to estimate $H(z)$ and the diameter-angular distance $D(z)$. Specifically, if $\mu_r$ is the value obtained assuming a particular arbitrary reference model, then $\mu$ depends on the true cosmological model as  $\mu=\mu_r H/(H_r \alpha)$ and $k$ as $k=\alpha k_r$, where \cite{2003ApJ...598..720S,Amendola:2004be}
\begin{equation}
    \alpha=\frac{\sqrt{H^2 D^2 \mu_r^2-H_r^2D_r^2(\mu_r^2-1)}}{H_r D}
\end{equation} 
This of course renders the results, in general, dependent on the background model. On the other hand, $H$ and $D$ can be estimated from supernovae and from the scale of the baryon acoustic oscillations independently of the cosmological expansion. Therefore, provided one can achieve precise constraints on $H$ and $D$ at the relevant redshifts, then $\mu,k$ can be determined in a model-independent way, and the arguments of this paper remain valid. If instead $H$ and $D$ are not well-measured, then our treatment remains valid only if one replaces $\beta$ and $P$ with the generalized model-independent observables $\beta (H/\alpha)^2$ and $P(k=\alpha k_r)$, respectively.

\section{Fisher matrix for two tracers}

Now we move to the case of two tracers with different biases, so we define $B_{i}=1+\beta_{i}\mu^{2}$, with $i=1,2$. The data covariance is identical to Eq. (\ref{Eq:correlation}), with the replacement $P_i \to B_i^2 P_i$, and with $N_{i} \to 1 + 1/(B_{i}^2P_{i})$, which can then be used in Eqs. (\ref{Eq:Fishbar2})-(\ref{Eq:Fishbar3}) to derive the Fisher matrix density for two tracers. Alternatively, we can use directly Eq. (\ref{Eq:MTFish}), with $P_i \to B_i^2 P_i$, arriving at the same expression.

Notice that the determinant of this Fisher matrix is zero by construction, and the reason is that for each value of $k$ and $\mu$ there is a complete degeneracy between the amplitude of the power spectra, $P_i$, and the redshift distortion parameters $B_i$. However, by integrating over $\mu$ we are in effect summing the Fisher matrices for the different values of $\mu$, which is what allow us to obtain independent constraints for the power spectrum as well as for $\beta_i$. In practice, this means combining the multipoles ($\ell=$0, 2 and 4) of the redshift-space power spectrum to extract independent constraints for those quantities. 

So far we have assumed that $P_1$ and $P_2$ are independent. This might be the case in some applications, but is too general for our scope. In fact, in our case the two power spectra are just biased  versions of the same underlying dark matter distribution, $P_{1,2}=\bar{n}_{1,2}b_{1,2}^2\langle\delta_m^2\rangle$. They are therefore related as
\begin{equation}
    P_2=P_1\frac{\beta^2_1}{\beta^2_2}q
\end{equation}
where $q=\bar{n}_2/\bar{n}_1$.

We could now replace everywhere $P_2$ with $P_1$ and reduce our degrees of freedom from four to three parameters. However, we are interested in comparing the results of the scenario with two tracers with the alternative where we combine both of them into a single one. In that case it is more convenient to replace both $P_1$ and $P_2$ by the resulting total spectrum $P$ as new parameter. The total spectrum  can be obtained as follows: first, consider that the counts of a single (combined) tracer are related to the counts of the two distinct tracers by $n_{t} (\vec{x}) = n_1 (\vec{x}) + n_2 (\vec{x})$,
which leads, through the spatial mean, to $\bar{n} = \bar{n}_1 + \bar{n}_2$. Then, using the definition of ``local bias'', $\delta_1 = \delta n_1/\bar{n}_1 = b_1 \delta_m$, we obtain the bias of the single tracer as:
\begin{equation}\label{eq:comp-b2}
    b=\frac{\bar n_1 b_1+\bar n_2b_2}{\bar n_1+\bar n_2} =\frac{ b_1+q b_2}{1+q}
\end{equation}
so that the density contrast obeys the relation
\begin{equation}
\label{eq:comp-b3}
    \bar{n} b\delta_m =\frac{ b_1+q b_2}{1+q} \, \bar{n} \delta_m \; .
\end{equation}

From Eq. (\ref{eq:comp-b3}) we recognize that the case of a single (combined) tracer can be related to the case of two tracers via the variances of the Fourier transforms of the density fields:
\begin{eqnarray}
\label{eq:replace1}
    \sqrt{\bar{n}_1 P_1} &=& \sqrt{\bar{n} \, P} \cos^2 \phi \, , \\
\label{eq:replace2}
    \sqrt{\bar{n}_2 P_2} &=& \sqrt{\bar{n} \, P} \sin^2 \phi \, ,
\end{eqnarray}
where
\begin{eqnarray}
\cos^2 \phi &=& \frac{1}{1+Y} \, , \\
\sin^2 \phi &=& \frac{Y}{1+Y} \, ,
\end{eqnarray}
with $Y  = q \, \beta_1/\beta_2$. In particular, we can write:
\begin{equation}
    P_1 + P_2 = P \times \frac{(1+q)(1+Y^2/q)}{ (1+Y)^2} \, ,
\end{equation}
where it is clear that the factor in right-hand-side reduces to 1 when $\beta_1 = \beta_2$, since in that case $Y=q$. Indeed, in that particular case Eqs. (\ref{eq:replace1})-(\ref{eq:replace2})  reduce to $\sqrt{P_1} \to \sqrt{P} \cos \phi$ and $\sqrt{P_2} \to \sqrt{P} \sin \phi$, from where it follows that $P_1 + P_2 \to P$.

These expressions can be generalized to $N$ tracers in terms of spherical coordinates: the variance in the single-tracer case, $\sqrt{\bar{n} P}$, becomes the (square of the) radial coordinate, while the variances of the original tracers are the projections of that radial coordinate into the different axes, according to the angle variables in an $N$-dimensional spherical coordinate system.

Therefore, using Eqs. (\ref{eq:replace1})-(\ref{eq:replace2}), we can replace $P_{1,2}$ in favour of the total spectrum $P$,
and use as parameters the reduced set $X=\{\log P,\log\beta_1,\log\beta_2\}$, obtaining the matrix $\bar F^{\mu}_{\alpha\beta}$, to be averaged over $\mu$. 
Although the ratio $q=\bar{n}_2/\bar{n}_1$ is in principle independent of $\beta_{1,2}$, highly biased populations are expected to be sparser, so small $\beta$ often implies small $\bar{n}$. For this reason, in the plots and tables below we assume as an illustrative case that $\bar{n}_i={\rm const}\cdot \beta_i^{2}$, so that $P_1=P_2$. However, the analysis is general.

Averaging over $\mu$ we have the Fisher matrix per unit phase-space volume
\begin{equation}
\bar{F}_{\alpha\beta}=\frac{1}{2}\int_{-1}^{+1}d\mu\bar{F}^{(\mu)}_{\alpha\beta} \; ,
\end{equation}
where $\alpha,\beta=1,2,3$. Since $\bar F _{\alpha\beta}^{(\mu)}$ are rational functions, the integrals are analytical, but extremely cumbersome, so here we  display only the numerical results and make the numerical code publicly available\footnote{{\it Mathematica} notebook at the link github.com/itpamendola/multipletracers.}. The fully marginalized
relative errors on $\beta_{i}$ are
\begin{equation}
\sigma_{\beta_{1}}^{2}=(\bar{F}^{-1})_{22} \quad , \quad \sigma_{\beta_{2}}^{2}=(\bar{F}^{-1})_{33} \; ,
\label{eq:sigmabetaa}
\end{equation}
and are functions of the fiducial values $P,\beta_1,\beta_2$ and of $q=\bar{n}_2/\bar{n}_1$. Obviously, swapping the values of $\beta_1,\beta_2$, and for $q\to 1/q$,  one has $\sigma_{\beta_1}\leftrightarrow \sigma_{\beta_2}$. 
  One could also consider the conditional (or maximised) relative error \begin{equation}
(\sigma^c_{\beta_{1}})^{2}=(\bar{F}_{22})^{-1} \; ,
\end{equation}
which gives the best possible estimate of $\beta_1$, achieved when all the other variables are perfectly measured (and analogously for $\beta_2$). Fixing some of the parameters does not imply that we have infinite signal-to-noise for $P$  (in which case we would have a perfect measurement of the $\beta$'s as well), but that other data can help constrain some of the parameters, helping to break degeneracies. Hence, the conditional errors are useful as a limit that can be reached as auxiliary data sets are included.
However, in the following we focus on the more conservative marginalized error.

In view of the comments in Sect. III, it is important to notice that $\bar{F}$ is not diagonal and that, in particular, the correlations $\sigma_{P\beta_1},\sigma_{P\beta_2}$ do not vanish for a finite SNR. Therefore, any parametrization of $P$ will influence the estimation of the $\beta$ parameters. This is why we need to take the waveband $P(k)$ itself as parameter: being a directly observable quantity, one does not need to introduce any model to estimate it. The quantities $\beta_{1,2}$ and $P(k)$, although not statistically independent, are therefore direct combinations of the data, i.e., they are model independent statistics. In the limit of high SNR, they also become statistically independent and therefore, in this limit, the cosmic variance uncertainty does not propagate from $P$ to $\beta_{1,2}$.

  In order to perform a fair comparison between the single- and two-tracer case, we take $b$ as in (\ref{eq:comp-b2}), from which we get the combined $\beta$:
\begin{equation}\label{eq:comp-b}
    \beta=\frac{ (1+q)\beta_1\beta_2 }{\beta_2+q \beta_1} \; ,
\end{equation}
whose relative uncertainty can be obtained by propagating the 
covariance of $\{ \log \beta_1 , \log \beta_2 \}$, with the result:
\begin{equation}
\label{eq:varbs}
    \sigma_{\beta}^2=
    \frac{\beta_2^2\sigma_{\beta_1}^2+q^2\beta_1^2\sigma_{\beta_2}^2+2q\beta_1\beta_2\sigma_{\beta_1\beta_2}^2}{(\beta_2+\beta_1 q)^2} \; .
\end{equation}
It is this quantity that is compared to the variance (\ref{eq:errbeta}) for a single tracer in all our numerical results.

\begin{figure}
\includegraphics[width=12cm]{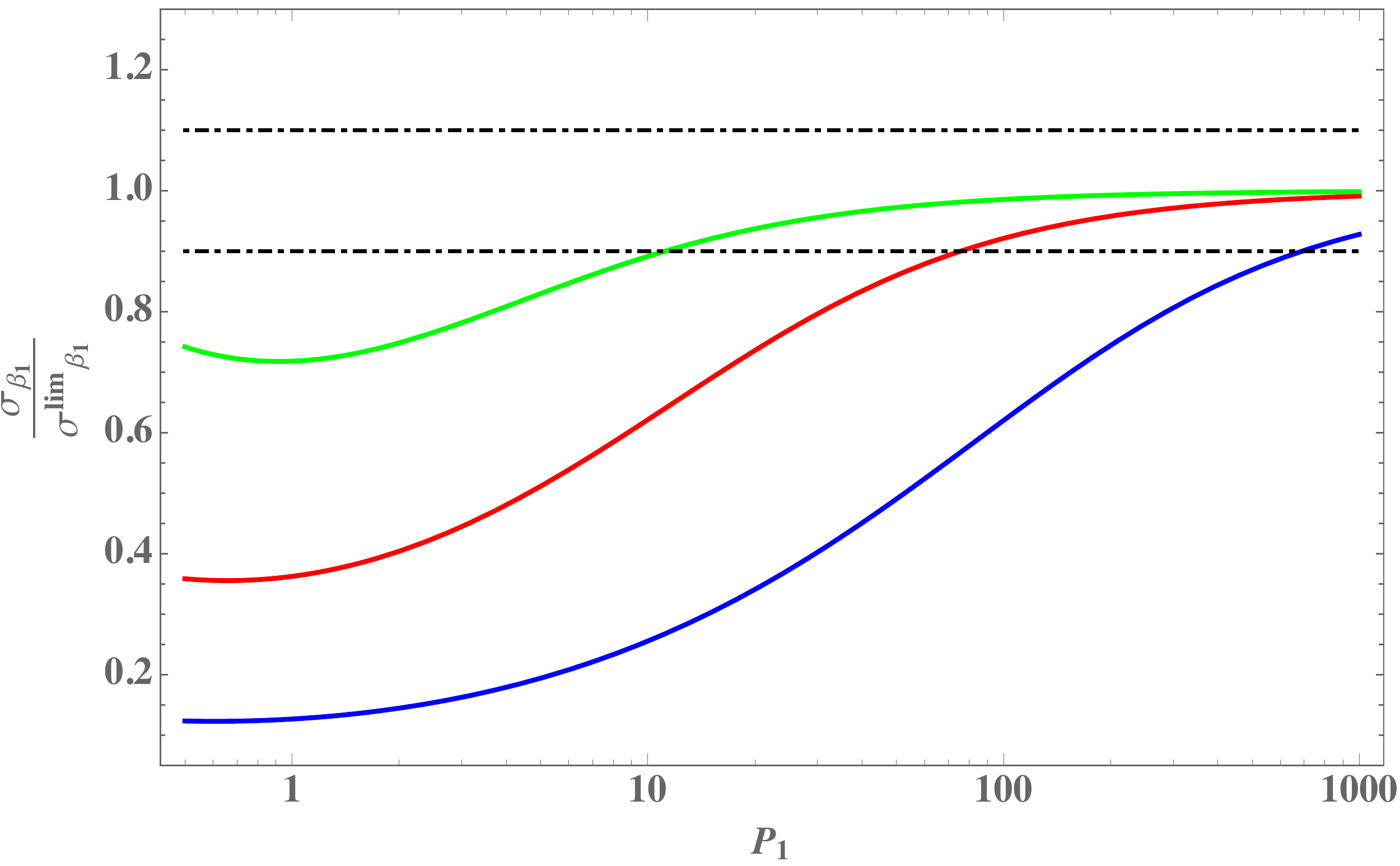}
\caption{\label{fig:compare-limit}   Ratio of the exact result and the $P\gg 1$ approximation of Eq. (\ref{eq:siglim})
  for $\beta_{2}=1$ and  $\beta_1=0.1$ (top curve, green in the color version), 0.5 (middle curve, red), 0.8 (bottom curve, blue). The dot-dashed horizontal lines mark the $10\%$ level.   }
\end{figure}

The limit for large $P$  can be obtained analytically if $\beta_1\not=\beta_2$,
\begin{equation}
    (\sigma^{{\rm lim}}_{\beta_{1}})^{2}=\frac{1}{P\Delta}\frac{\sqrt{\beta _1} \left(\beta _2+\beta _1 q\right){}^2 \left(\Delta 
   \sqrt{q} \left(T \left(z_a^*\right){}^{3/2} z^*_b-T^*
   z_a^{3/2} z_b\right)+2 i \sqrt{\beta _1\beta_2}  \sqrt{q+1} |z_b|^2
   \right)}{2 \sqrt{\beta _1} \beta _2^2   \sqrt{q} (q+1)
   \left(T  \sqrt{z_a^*}z_a z^*_b-T^* \sqrt{z_a}
   z_a^* z_b\right)-4 i \beta _2^{3/2} \Delta   q
   \sqrt{q+1} T T^* |z_a| } \; ,
   \label{eq:siglim}
\end{equation}
where:
\begin{align}
T & =\tan ^{-1}\left(\frac{\sqrt{q+1} \sqrt{\beta _1\beta_2} }{\sqrt{z_a} z_b^*}\right) \; , \\
z_a&=\sqrt{q}+i  \; , \\
z_b&=\sqrt{\beta_1 \sqrt{q}+i\beta_2}  \; , \\
\Delta&=\beta_1-\beta_2 \; .
\end{align}
Here we see very clearly the power of the multiple-tracer method: while the single-tracer variance of $\beta$ reaches a constant for large $P$ (see Eq. \ref{eq:st-largep}), for two tracers it decreases as $1/P$, and becomes smaller for larger $\Delta$. The same $(P\Delta)^{-1}$ trend applies to $\sigma^2_{\beta_2}$ and to $\sigma^2_{\beta_1\beta_2}$, and therefore to $\sigma^2_\beta$ as well. As can be seen in Fig. (\ref{fig:compare-limit}), for $\beta_2=1$ this asymptotic expression performs better than $10\%$ for any $P>30$ and $\beta_1<0.4$.

\begin{figure}
\includegraphics[width=8cm]{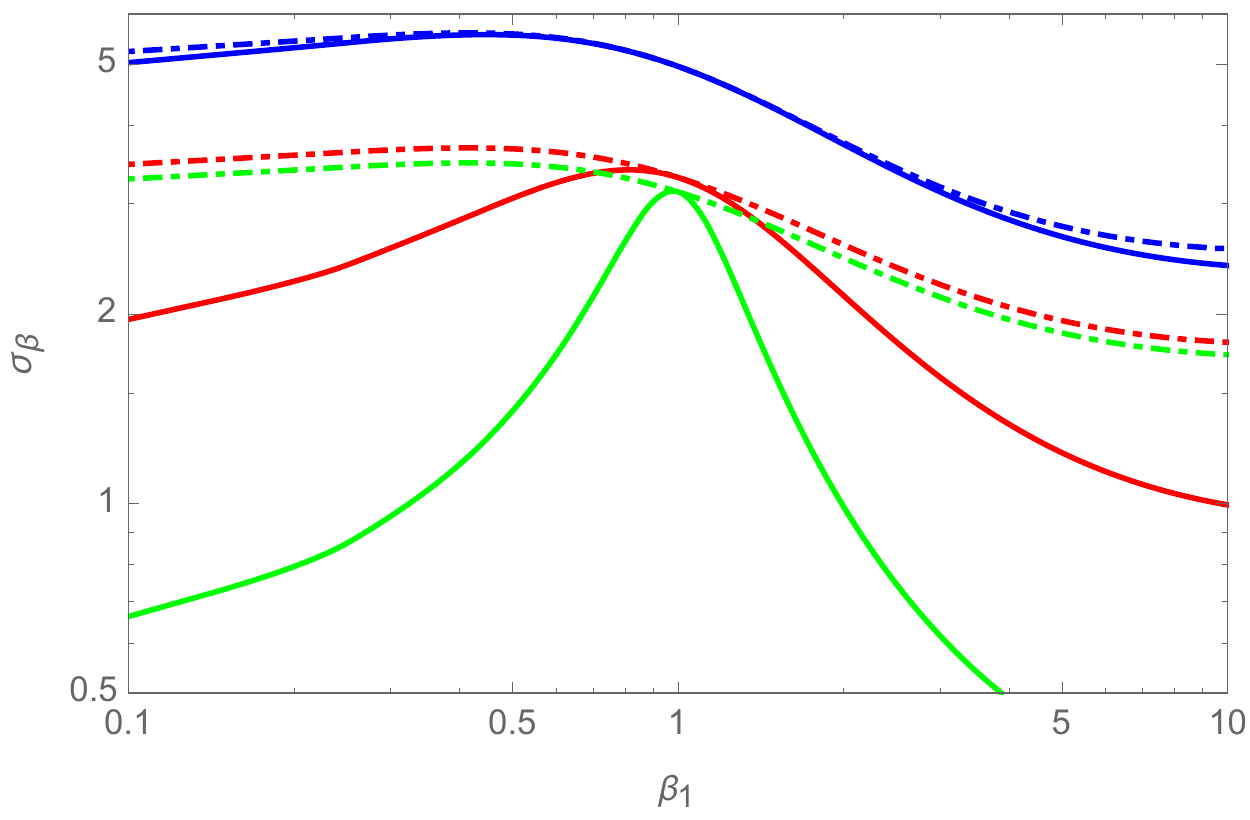}
\hskip 0.5cm
\includegraphics[width=8cm]{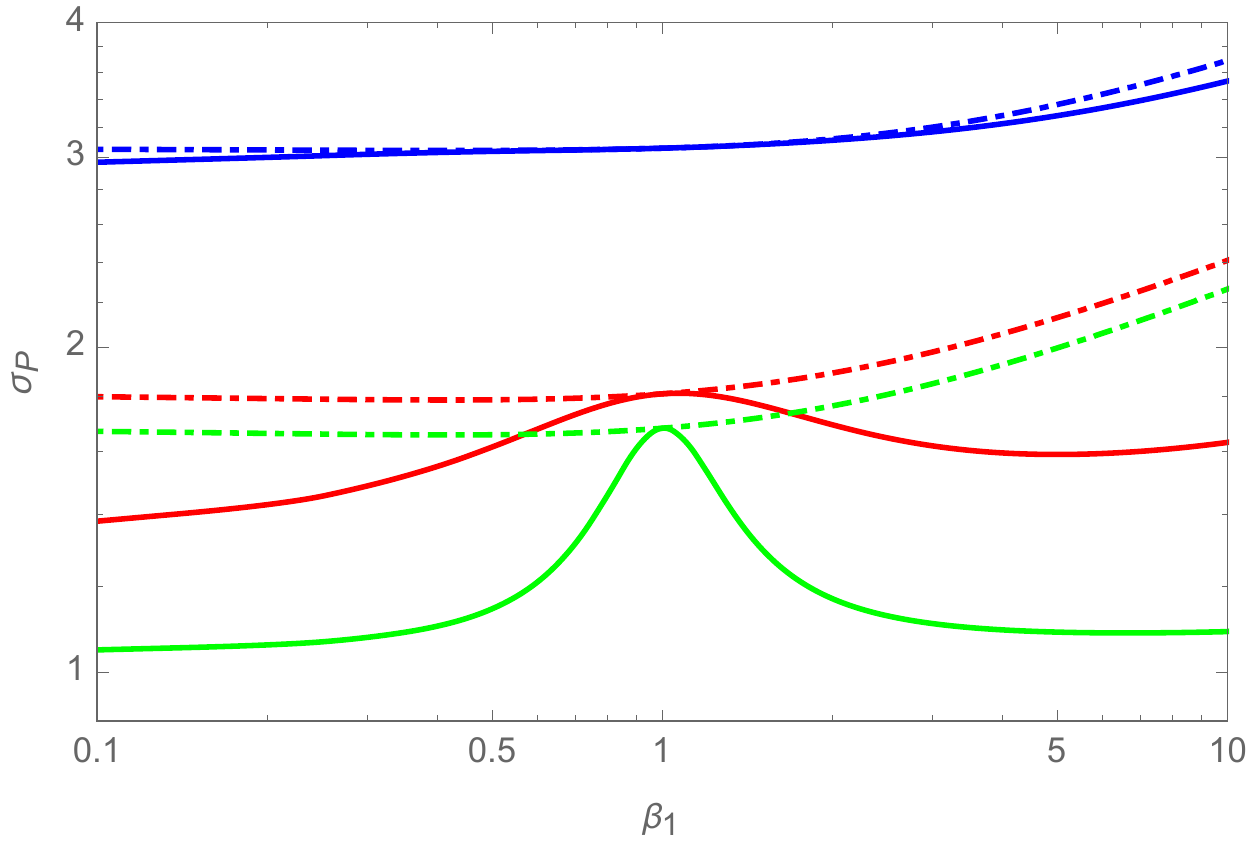}
\caption{\label{fig:errorsbp_b} 
Marginalized relative errors $\sigma_{\beta}$ (left) and 
$\sigma_P$ (right) as a function of $\beta_{1}$. 
Here we fixed $\beta_2=1.0$ and took the values $P=1$ (top curves, blue in color version), 10 (middle curve, red), 
and 100 (bottom curve, green), and assumed that the number densities of the tracers scale as $\bar{n}_i \propto \beta_i^2$.
The full curves correspond to the two-tracer 
marginalized relative error of the mean $\beta$, as given by Eq. (\ref{eq:varbs});
the dashed curves correspond to the single-tracer case, where we combined the two tracers into a single one. 
As expected, for $\beta_1=\beta_2=1$ the single- and 
two-tracer cases coincide.
Here and in the following plot, the errors should be multiplied by the appropriate phase space volume factor $\gamma $ -- see Tables I and II for the case of the Euclid survey.}
\end{figure}

\begin{figure}
\includegraphics[width=8cm]{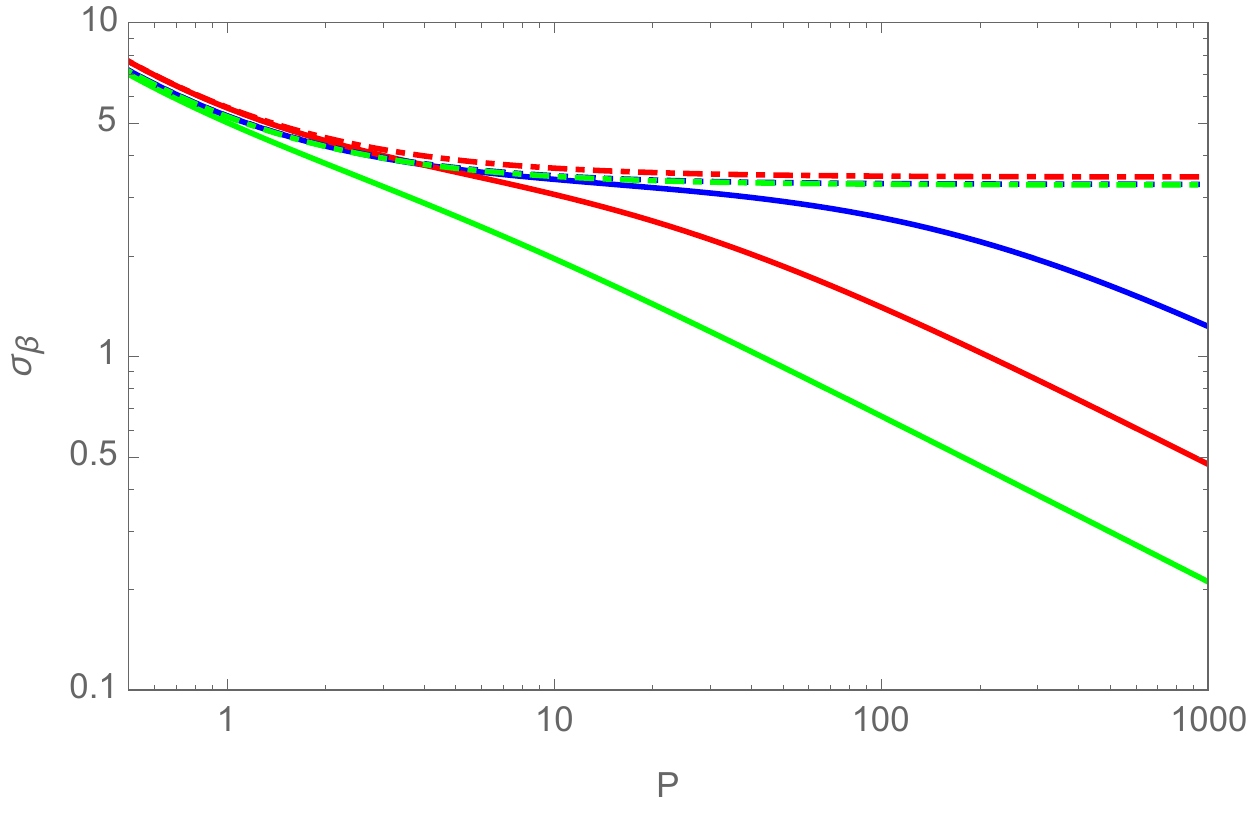}
\hskip 0.5cm
\includegraphics[width=8cm]{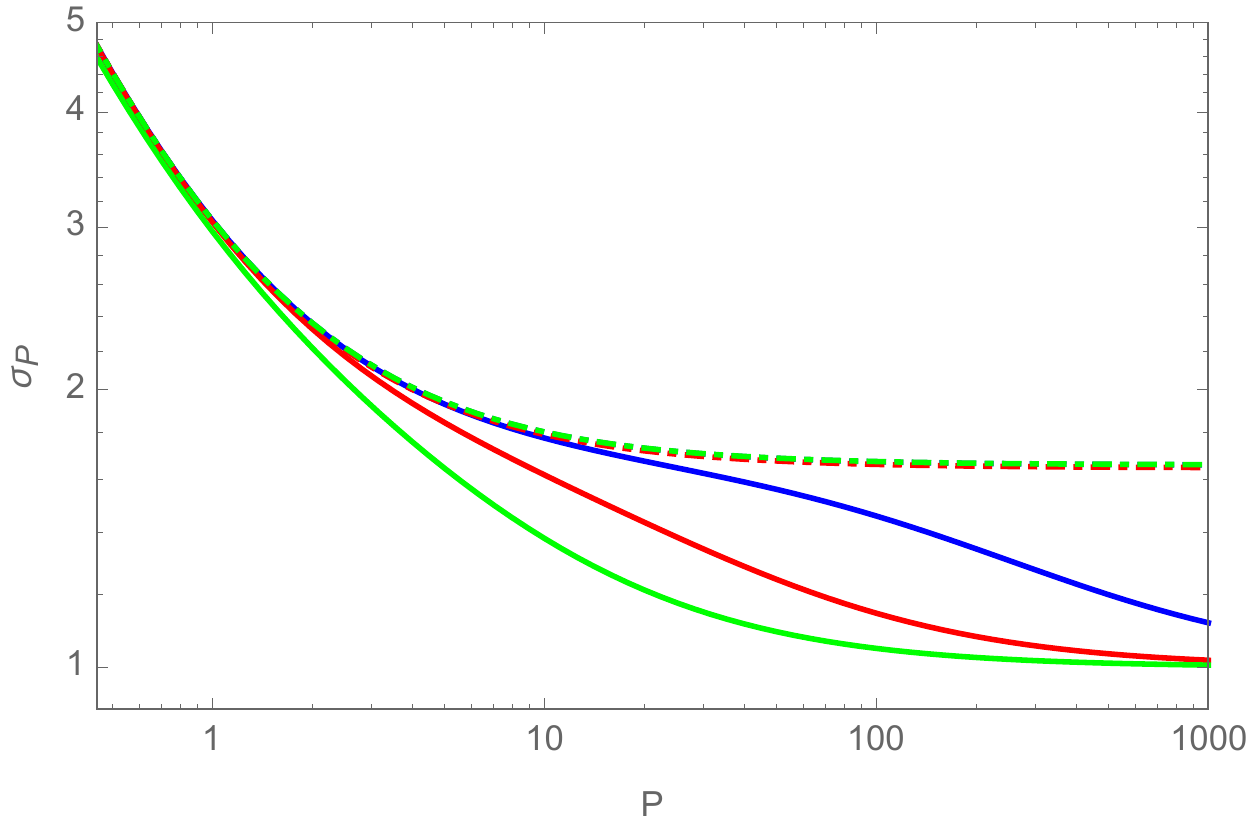}
\caption{\label{fig:errorsbp_p} Marginalized relative errors $\sigma_{\beta}$ (left) and 
$\sigma_P$ (right) as a function of $P$. 
Here we fixed $\beta_2=1.0$ and took the values $\beta_1=0.8$ (top curves, blue in color version), 0.5 (middle curve, red), 
and 0.1 (bottom curve, green). As in the previous figure, the number densities of the tracers are assumed to scale as $\bar{n}_i \propto \beta_i^2$. The full curves correspond to the two-tracer 
marginalized relative error, and the dashed curves correspond to the single-tracer case.}
\end{figure}

We report in figures (\ref{fig:errorsbp_b}) and (\ref{fig:errorsbp_p}) the relative marginalized errors on $\beta$ and on $P$ as a function of $\beta_{1}$ and as a function of $P$, respectively.
When the two tracers are used in the analysis we denote the results in solid lines, and when a single-tracer is used the results are shown by the dot-dashed lines.
For the plots as a function of $\beta_1$ we fixed $\beta_2=1.0$ and used three values for $P$: 1, 10 and 100.
For the plots as a function of $P$ we again fixed $\beta_2=1.0$ and used three values for $\beta_1$: 0.8, 0.5 and 0.1. As one can see from Fig. (\ref{fig:errorsbp_b}), the two-tracer error can be significantly better than the single-tracer one, and the advantage increases sharply for  $P\gg 1$.
These figures represent our main result.

As we can see both from Fig. (\ref{fig:errorsbp_p}), as well as from Eq. (\ref{eq:f11}), the relative error on $P$ reaches a constant value for large $P$, contrary to $\beta$, which expresses the well-known fact that the multi-tracer technique does not cancel cosmic variance for the observable $P$ -- see, e.g., \cite{Abramo:2013}. However, as Fig.  (\ref{fig:errorsbp_p}) shows, with the inclusion of redshift-space distortions  one can improve the measurement of $P$ in the two-tracer case compared with a single tracer
for any $P$ larger than $\sim {\cal{O}}(1)$. For large $P$, in fact, the two-tracer limit is exactly 1, while for a single tracer we have the expression Eq. (\ref{eq:st-largepp}), which yields values around 1.5-1.6 in the range $\beta\in (0.1-1)$.

\begin{figure}
\includegraphics[width=10cm]{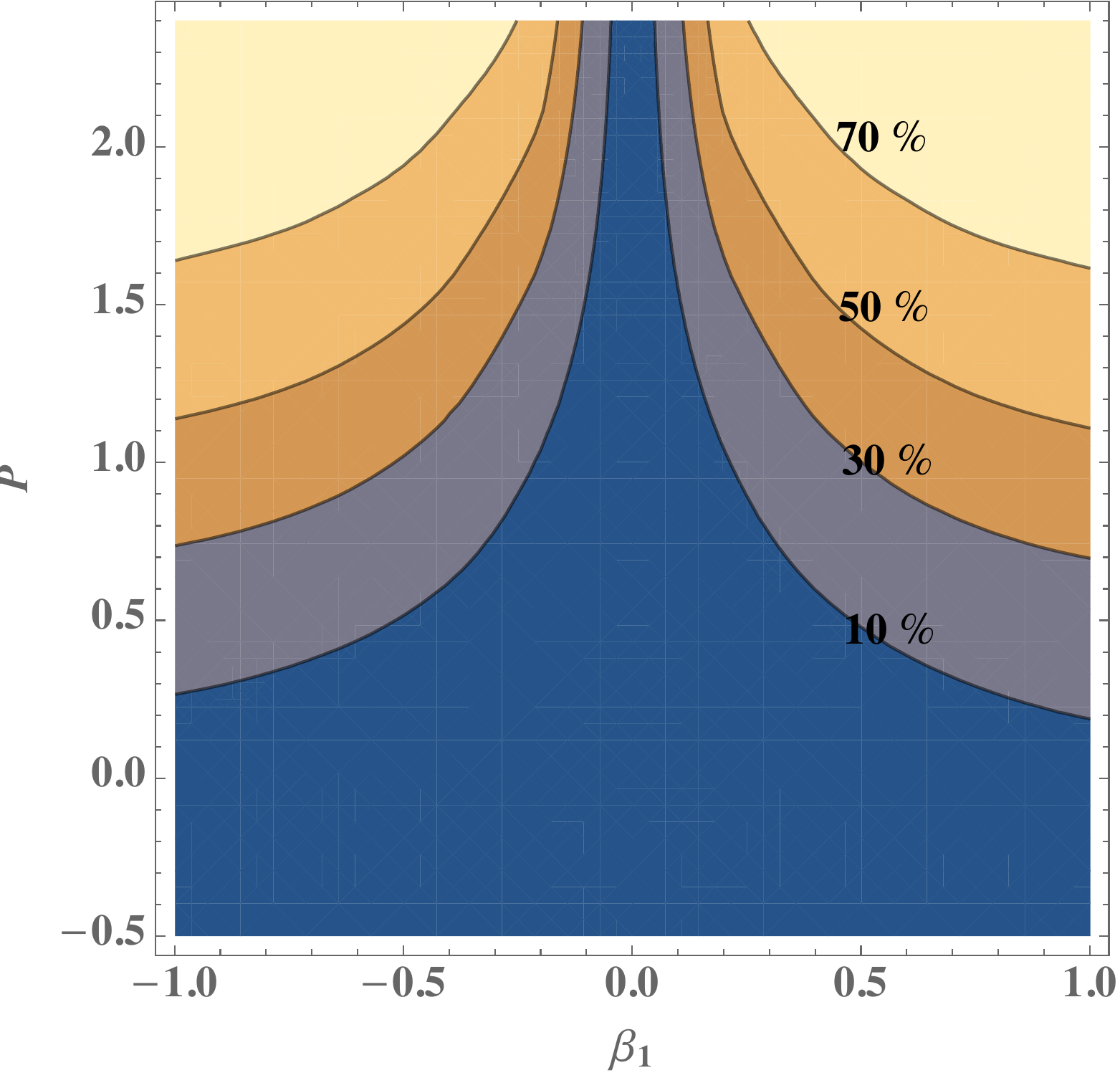}
\caption{\label{fig:cplot}   Contour plot of the error gain, $\delta_\sigma=\sigma_\beta({\rm 2\, tracers})/\sigma_\beta({\rm 1\, tracer})-1$, labelled by the percent gain (e.g., $10\%$ means $\delta_\sigma=-0.1$).  We assume $\beta_2=1$.}
\end{figure}

In Fig. (\ref{fig:cplot}), finally, we display the regions in the space $(\log_{10} P,\log_{10}\beta_1)$ in which the relative marginalized error for the combined tracers is smaller than the corresponding single-tracer case by the indicated percentage. As is already clear from the previous plots, the multi-tracer gain increases with larger $P$, as well as for larger differences in $\beta$'s 
 -- in fact, in addition to halos or galaxies, including voids (which have negative bias) further improves the multi-tracer gains \cite{2018arXiv181204024C}. It is instructive to compare our results for the relative gain of the multi-tracer analysis
in a model-independent way with those of, e.g., Ref. \cite{Hamaus:2012ap}, who first highlighted the advantage
of this method for measuring the matter growth rate, but assumed an underlying model ($\Lambda$CDM).
Our results should also be compared with those obtained for the GAMA survey \cite{2013MNRAS.436.3089B}, in particular Fig. 14 of that paper. The final result of Ref. \cite{2013MNRAS.436.3089B} was an improvement of $\sim$ 10 -- 20 \% for that data set using the multi-tracer analysis, but the Fisher forecast of that paper is consistent with our findings --- even if in our case we did not have to assume a cosmological model.

 From the numerical and analytical formulae, we can draw several interesting conclusions:
\begin{itemize}
\item  For our choice of $q$, the two-tracer error on the combined $\beta$ is  smaller than the single-tracer
one  for all $P$  (which in our notation is the signal-to-noise ratio). The advantage becomes significant for $P \gtrsim 10$. For instance, if $\beta_1=0.1,\beta_2=1$, the error is halved (with respect to single tracer)  for $P \gtrsim 10$, close to the typical Euclid value (see Table \ref{tab:An-approximation-to}). 
\item We explored the full parameter space $\log P,\log\beta_1,\log\beta_2 $ and $q$ within the range $10^{-2},10^2$ with $10^5$ random points, and in all cases we found a positive gain of the two-tracer method versus the corresponding one-tracer one; we conjecture therefore that the two-tracer method is indeed always advantageous.
\item The single-tracer error on $\beta$ reaches a constant value for
$P\to\infty$, see Eq. (\ref{eq:st-largep}). For the two-tracer case, $\sigma_{\beta} \to P^{-1/2}$
for $P\to\infty$: this shows how multiple tracers beat
cosmic variance.
\item For all realistic cases [$P_1\in(1-100)$ and $\beta_{1}\approx\beta_{2}\in(0.1-1)$]
the relative error on $\beta$ per phase-space unit is of order unity (more exactly, between 1 and 5). For Euclid, one should multiply the errors by the $\gamma $ factors listed in Tab. \ref{tab:An-approximation-to}.  As an example,  Tab. \ref{tab:An-approximation-to-2} gives the exact values for a realistic choice of bias and compares them with the single-tracer case.
\item The error for $P$ reaches a constant value for large $P$ both for the single- and the two-tracer case. The two-tracer asymptotic value is however roughly $60\%$ smaller, see Fig. (\ref{fig:errorsbp_p}), right panel.
\item More results can be obtained by running the publicly available code (see footnote 2) which gives the relative errors per unit phase-space on $P,\beta_1,\beta_2$ for one and two tracers, combined or separate, and for any number density ratio $q$.
\end{itemize}

\begin{table}
\[
\begin{array}{cccccc}
 &  & \multicolumn{2}{c}{one\; tracer}  
 &
\multicolumn{2}{c}{two\; tracers} 
\\
z_1 & z_2 &\gamma(k_1) \sigma_{\beta} &
\gamma(k_2) \sigma_{\beta} &
\gamma(k_1) \sigma_{\beta} &
\gamma(k_2) \sigma_{\beta}\\
\hline
 0.7 & 0.9 & 0.6 & 0.12 & 0.29 & 0.058 \\
 0.9 & 1.1 & 0.57 (0.57-0.58) & 0.11 (0.11-0.12) & 0.27 (0.29-0.35) & 0.053 (0.057-0.070) \\
 1.1 & 1.3 & 0.57 (0.57-0.58) & 0.11 (0.11-0.12) & 0.27 (0.33-0.41) & 0.054 (0.066-0.081) \\
 1.3 & 1.5 & 0.57 (0.58-0.60) & 0.12 (0.12-0.12) & 0.29 (0.39-0.48) & 0.058 (0.078-0.095) \\
 1.5 & 2.1 & 0.43  & 0.087 & 0.28  & 0.057 \\
\end{array}
\]
\caption{\label{tab:An-approximation-to-2}
Forecasts of errors for a Euclid-like survey, same specifications as in Tab. I. In each redshift bin, we  fix $\beta_1=f(z)/b(z)$, where $f(z)\approx\Omega_m^{0.54}$ is the $\Lambda$CDM growth rate and $b(z)\approx 0.7z+0.7$, as in \cite{Merson:2019vfr}, Table 4, WISP calibration, and $\beta_2=\beta_1/2$. Notice that the power spectra listed in Tab. I are now multiplied by $b(z)^2$.  First two columns: redshift bins. Third and fourth column:  relative errors for $\beta$ for a single tracer,   at $k_{1}=0.01$ $h$ Mpc$^{-1}$ and $k_{2}=0.05$
$h$ Mpc$^{-1}$. Last two columns: same for the two-tracer case.   }
\end{table}

\section{Projecting over parameters}

So far we have been trying to be as model independent as possible. In practice, one has often a model with a finite number of parameters, so one needs to project the Fisher matrix onto the parameter set. Suppose for instance that $\beta_{1,2}$ are independent of $k$ and parametrized by  a number of parameters $p_{\bar i}$ (for simplicity, we assume the same parametrization for both $\beta$'s), $\beta_{1,2}=\beta_{1,2}(z;p_{\bar i})$. Then the Fisher matrix densities for the shell at redshift $z$ can be summed over the $k$ bins
\begin{equation}
    F_{\alpha\beta}(z)=\sum_j \gamma^{-2}(z,k_j) \bar F_{\alpha\beta} \; .
\end{equation}
Defining now as $\theta_\alpha=(\log P, \log\beta_1,\log\beta_2)$ the old set of parameters, and by  $\theta_{\bar\alpha}=(\log P, p_{\bar i})$ the new one, we project over
 $\theta_{\bar\alpha}$  and finally sum over redshift slices $z_i$, and obtain,
\begin{equation}
     F_{\bar\alpha \bar\beta}= \sum_{i} \sum_{\sigma \tau} J(z_i)_{\sigma\bar\alpha} F(z_i)_{\sigma\tau} J(z_i)_{\tau \bar\beta }
\end{equation}
where the Jacobian is:
\begin{equation}
    J(z_i)_{\alpha\bar \sigma}=\frac{\partial \theta_\alpha}{\partial \theta_{\bar\sigma}}\big |_{z=z_i}=
    \begin{bmatrix}
   1       & 0 & 0 & 0 &\ldots \\
    0        & \frac{\partial\beta_1}{\partial p_1} & \frac{\partial\beta_1}{\partial p_2} & \ldots \\
    0        & \frac{\partial\beta_2}{\partial p_1} & \frac{\partial\beta_2}{\partial p_2} &\ldots 
\end{bmatrix}_{z=z_i}
\end{equation}
In this way the parameters $p_{\bar i}$ can be constrained much more stringently than the values of $\beta_{1,2}$ in any given $(z,k)$-bin. The resulting constraints will now obviously depend on the chosen $\beta$  parametrization, but would still be independent of the cosmological model.

\section{Conclusions}

We have shown that using two or more tracers of large-scale structure (e.g., galaxies of different types) it is possible to measure the redshift distortion parameter $\beta$ in each $k$- and redshift-bin in a model-independent way, with an accuracy that is not limited by cosmic variance. 

Here, model independent means that the constraints on the band power spectra $P_{1,2}$ and on the redshift distortion parameters $\beta_{1,2}$ do not depend on the background expansion rate\footnote{At least as long as one has a good estimation of the Hubble function and the angular-diameter distance through supernovae and BAO, as we discussed at the end of Sect. III.}, nor on the evolution of perturbations, nor on the initial conditions, and therefore apply to any cosmological model in the linear regime.  We have found that if the SNR is much larger than unity (of the order of 10 or more, depending on $\beta$), the $\beta$  parameters for two tracers can be estimated with significantly more accuracy compared with the case of a single (combined) tracer, thereby allowing an accurate estimate of the bias ratio $b_1/b_2$ for two species. Based on extensive numerical evidence, we conjecture that the two-tracer approach is always more constraining than the single-tracer one.
Numerical results for any  combination of parameters can be easily obtained by running a publicly available code (see footnote 2).

 Although our computations were performed in the context of the galaxy power spectrum in the flat-sky (or distant observer) approximation, where the effects of redshift distortions are encapsulated in the Kaiser term $(1+\beta\mu^2)^2$, the result should remain valid also in full-sky surveys such as Euclid  (as long as we stay in the linear regime), where the same signal would be found in the ratios of the angular power spectra of the different tracers.

In addition to the redshift distortion parameters $\beta$, by comparing the redshift-space distortion pattern of different tracers we can also  measure the velocity dispersion of galaxies in collapsed structures -- the ``Fingers-of-God'' effect \cite{1972MNRAS.156P...1J}. In particular, the scale-dependent signature of redshift distortions in the non-linear regime might be especially useful to disentangle these parameters from the shape of the spectrum as well as the growth rate in the linear regime. We are now exploring this new window into the small-scale clustering in redshift space.

\section*{Acknowledgments}

L. R. A. would like to thank CNPq and FAPESP (grant 2018/04683-9) for financial support; L.A. acknowledges support from DAAD to attend the II Workshop on Current Challenges in Cosmology, Bogotá (Colombia), where this work was started. 
The authors would also like to thank the organizers of the Workshop for creating an opportunity for  scientific discussions such as the ones which originated this paper. L.A. also thanks Anke Ackermann for checking the equations and spotting a few typos.

\bibliography{99}

\end{document}